\documentclass[preprint]{aastex}

\shorttitle{Canis Major Globular Clusters}
\shortauthors{Forbes, Strader \& Brodie}

\def\etal{{\it et al.~}}

\begin{document}

\title{The Globular Cluster System of the Canis Major Dwarf Galaxy}

\author{Duncan A. Forbes}
\affil{Centre for Astrophysics \& Supercomputing, Swinburne University,
  Hawthorn, VIC 3122, Australia}
\email{dforbes@swin.edu.au}

\and

\author{Jay Strader and Jean P. Brodie}
\affil{UCO/Lick Observatory, University of California,
 Santa Cruz, CA 95064, USA}
\email{strader, brodie@ucolick.org}


\begin{abstract}

Prompted by the discovery of the accreted Canis Major dwarf
galaxy and its associated globular cluster (GC) system (Martin
\etal ), we investigate the contribution of accreted GCs to
the Galactic system. The Canis Major GCs, and those associated
with the Sagittarius dwarf galaxy, exhibit a range of 
galactocentric radii, prograde and retrograde motions, and
horizontal branch morphologies, indicating that such properties
are of limited use in identifying accreted GCs. By contrast, we
find that the age-metallicity relation (AMR) of these dwarf
galaxies is distinct from that of the main Galactic GC
distribution at intermediate-to-high metallicities ([Fe/H] $\ga
-1.3$). The accretion of GCs with a distinct AMR would explain much of
the apparent age spread in the Galactic GC system.
The Canis Major and Sagittarius AMRs are similar
to those of other Local Group dwarf galaxies and are consistent
with a simple closed-box chemical enrichment model -- a further
indication that these GCs formed outside of the Milky Way. 
The Canis Major GCs all have smaller-than-average sizes for their
galactocentric distances, lending further support to their origin
outside of the Milky Way. 
Our findings
suggest that accretion of similar mass dwarfs does not appear to have played a major
role in building the stellar mass of the thick disk or bulge of
the Milky Way.

\end{abstract}

\keywords{globular clusters: general -- galaxies: individual (Canis Major dwarf) -- galaxies: star clusters}

\section{Introduction}

The seminal paper of Searle \& Zinn (1978) argued that the
formation of the Milky Way galaxy was clumpy and rather
chaotic. Furthermore, they suggested that some Galactic globular
clusters (GCs) may not have formed {\it in situ} but rather in
proto-galactic fragments or dwarf galaxies that were later
accreted.  Several workers have attempted to identify candidates
for these accreted GCs. Methods have included the retrograde
motion of some metal-poor GCs (Rodgers \& Paltoglou 1984), the
variation of horizontal branch (HB) morphology with metallicity
in the outer halo (Zinn 1993), the Oosterhoff class (van den
Bergh 1993), the size-perigalactic distance relation (van den
Bergh 1995) and associations in phase-space (Lynden-Bell \&
Lynden-Bell 1995).

This subject was given impetus with the discovery of the accreted
Sagittarius (Sgr) dwarf galaxy and its system of globular
clusters (Ibata, Gilmore \& Irwin 1995).  It has four clearly
identified GCs (Terzan 7, Terzan 8, Arp 2 and M54), with M54 as a
prime candidate for the nucleus of the disrupted dwarf (Layden \&
Sarajedini 2000). Comparing the phase-space distribution of GCs
with the expected orbital path of the Sgr dwarf, Bellazzini
\etal~(2003a) confirmed the status of these four GCs and identified a
number of other possible GCs. The closest two GCs to the orbital
path are Pal 12 and NGC 4147. Bellazzini \etal~(2003b) found
evidence for Sgr dwarf `tidal debris' (i.e. stripped stars) 
around these GCs further supporting
their association with the Sgr galaxy. For the third closest GC, Pal 5,
Odenkirchen \etal (2003) concluded that it was {\it
not} part of the Sgr dwarf as it is on a different orbital
plane. Thus, we have a sample of six GCs which can be confidently
associated with the Sgr dwarf. For a magnitude of $M_V = -13.8$,
this corresponds to a specific frequency of S$_N$ = 18,
comparable to that of Fornax dwarf (S$_N$ = 29; Forbes \etal~2000), the
highest S$_N$ galaxy in the Local Group.

Recently, Martin \etal (2003) presented evidence for a second disrupted
dwarf galaxy in the Milky Way
(Martin \etal 2003). The `Monoceros Ring', discovered
earlier by Newberg \etal (2002), is now thought to be the tidal stream
of this disrupted dwarf (whose nucleus lies in the direction of
Canis Major). Crane \etal (2003) and Frinchaboy \etal (2004) 
compared the phase-space distribution of Galactic clusters with 
M giant stars in the Monoceros ring. Assuming a typical thick disk velocity dispersion
, they identified five GCs 
candidates of the disrupted dwarf. They are  NGC~2298,
NGC~2808, NGC~5286, Pal~1 and BH~176 respectively.
 
In the Canis Major discovery paper, Martin \etal~(2003) included
an N-body simulation of the dwarf's orbit over the last 2
Gyrs. They simulated both a prograde and retrograde orbit. Although
the prograde orbit is favoured, the retrograde orbit could not be
ruled out. Further observational constraints and a non-static
Galactic potential in the model should help to confirm which
orbit is
correct. Several GCs were found to have phase-space distributions
in common with both models, which led Martin \etal (2003) to
conclude they were associated with the Canis Major dwarf. They
are: NGC 1851, NGC 1904, NGC 2298 and NGC 2808. 
For an assumed magnitude similar to the Sgr dwarf
(Martin \etal~2003), the four GCs translate into a specific frequency of
S$_N$ = 12.  
In a follow-up paper, Bellazzini \etal
(2003c) argued that the spatial position and stellar populations of
the old open clusters AM-2 and Tom~2 imply that they too are
associated with the Canis Major dwarf. 

Here we derive an updated age-metallicity relation for the Sgr dwarf galaxy
and compare it to the Canis Major and Monoceros ring cluster
systems. We show that the Canis Major GCs reveal well-defined
age-metallicity and size-galactocentric distance 
relations. Finally, we briefly discuss the possible thick
disk/bulge association of the Canis Major GCs and assess the
overall importance of accretion in building the stellar mass of the Milky
Way.

\section{The Data}

In Table 1 we list various properties of the GCs associated with
the Sgr and Canis Major dwarf galaxies. We also
include the three GC candidates, in addition to NGC~2298
and NGC~2808, put forward by Crane \etal (2003). We will refer to the latter
three as the `Mono ring GCs' but remind the reader that the ring
is a tidal stream of the Canis Major dwarf observed near the
Galactic anti-center. 
Most ages and metallicities are taken from the homogeneous
compilation of Salaris \& Weiss (2002). However, metallicities for 
Terzan 8, M54, NGC~4147,  NGC~5286 and Pal~1 come from Harris
(1996). The age of Terzan 8 is assumed to be old, i.e. 13 Gyrs
old with M54 1 Gyr younger at 12
Gyrs based on the study of Montegriffo \etal~(1998), for 
NGC~4147 and NGC~5286 we assume an old age (13.0 Gyrs) based on
the colour-magnitude diagrams of Friel \etal (1987) and 
Samus \etal (1995) respectively, for Pal~1 we take the age
derived by Rosenberg \etal (1998). We note that most of these age estimates are
based on the magnitude difference between the horizontal branch
and the main sequence turnoff, and thus may suffer from unknown
systematic errors due to the second parameter effect.  
Other properties are from the catalog of Harris (1996).

From Table 1 it can be seen that both the Sgr and Canis
Major/Mono ring GC
systems 
have a variety of HB morphologies, with different magnitude and metallicity
distributions. The GCs also have both prograde and
retrograde motions (Rodgers \& Paltoglou 1984). Thus such properties are probably of limited
use in separating {\it bona fide} Milky Way GCs from those that
have been accreted from disrupted dwarfs.

\begin{deluxetable}{llccrcc}
\tablecaption{Dwarf Galaxy Globular Clusters}
\tablehead{Name & Age & [Fe/H] & R$_{GC}$ & HB & $M_V$ & r$_{h}$\\
              & (Gyrs) & (dex) & (kpc) & & (mag) & (pc)}
\startdata
{\it Sagittarius}  & & & & & &\\
M54 &      $12.0\pm2.0$ & $-1.79$ & 19.2 & 0.75 & $-10.0$ & 3.83\\
Terzan 7 & $7.5\pm1.4$  & $-1.00$ & 16.0 & $-1.00$ & $-5.1$ & 6.56\\
Terzan 8 & $13.0\pm1.5$ & $-2.00$ & 19.1 & 1.00 & $-5.1$ & 7.57\\
Arp 2 &    $11.5\pm1.4$ & $-1.84$ & 21.4 & 0.86 & $-5.3$ & 15.9\\
Pal 12 &    $6.4\pm0.9$ & $-0.82$ & 15.9 & $-1.00$ & $-4.5$ & 7.12\\
NGC~4147 & $13.0\pm1.5$ & $-1.83$  & 21.3 & 0.55 & $-6.2$ & 2.40\\
{\it Canis Major} & & & & & &\\
NGC~1851 & $9.1\pm1.1$ & $-1.23$ & 16.7 & $-0.36$ & $-8.3$ & 1.83\\
NGC~1904 & $12.6\pm1.3$ & $-1.67$ & 18.8 & 0.89 & $-7.9$ & 3.01\\
NGC~2298 & $12.9\pm1.4$ & $-1.85$ & 15.7 & 0.93 & $-6.3$ & 2.44\\
NGC~2808 & $10.2\pm1.1$ & $-1.36$ & 11.0 & $-0.49$ & $-9.4$ & 2.12\\
{\it Monoceros Ring} & & & & & &\\
Pal~1 & 8.0$\pm$2.0 & $-0.60$ & 17.0 & $-1.00$ & $-2.5$ & 2.15\\
NGC~5286 & 13.0$\pm$2.0 &   $-1.67$  & 8.40 & 0.86 & $-8.6$ & 2.21\\
BH~176 & 7.0$\pm$1.5  & 0.00  &   9.70 & \nodata & $-4.4$ & \nodata\\
\enddata
\end{deluxetable}

\section{The Age-Metallicity Relation of the Sagittarius Dwarf Galaxy}

The age-metallicity relation (AMR) of a stellar system provides a
key probe of the chemical enrichment history of that system
(e.g., Pagel \& Tautvaisiene 1995). Here, we derive an updated
age-metallicity relation for the Sgr dwarf (see e.g. Layden \&
Sarajedini 2000) using both the GC
system given in Table 1 and studies of field stars.

Ages and metallicities for three field star populations in Sgr were
measured by 
Layden \& Sarajedini (2000). Recently, Bonifacio \etal (2004)
found a very young and metal-rich stellar component. The mean
age-metallicity and range for these four stellar fields, and the six
GCs of the Sgr galaxy are shown in Fig. 1. 

\begin{figure}
\plotone{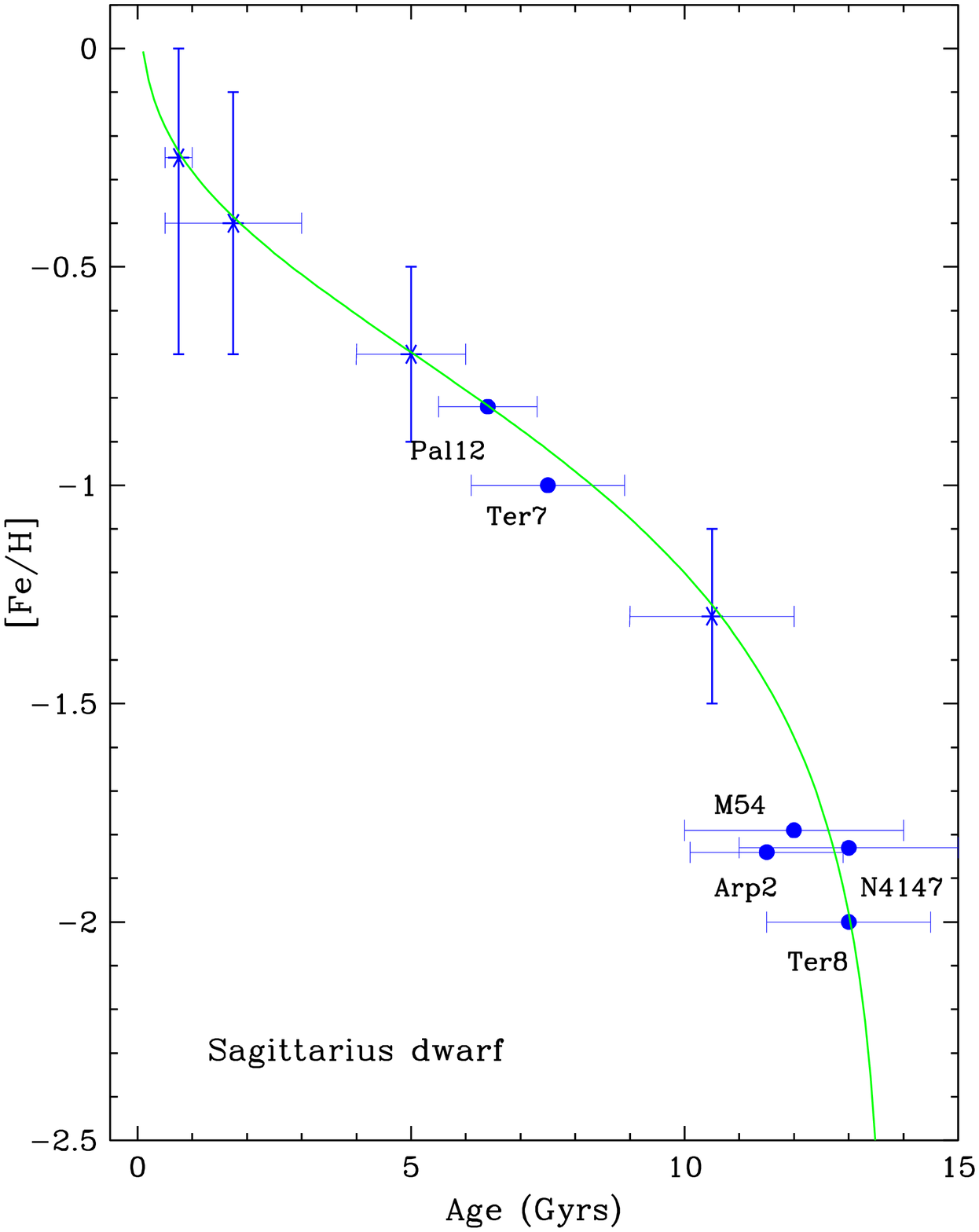}
\caption[forbes.f1.ps]{Age-metallicity relation for the Sagittarius
dwarf galaxy. The six 
globular clusters associated with Sgr dwarf are shown by filled
circles. The typical error in metallicity is $\pm$ 0.1 dex. 
The Sgr field star populations of Layden \& Sarajedini
(2000) and Bonifacio \etal (2004) are shown by 
asterisks.  The curve shows a simple closed-box age-metallicity
relation. It provides a good representation of the enrichment
history of the Sgr dwarf galaxy. }
\end{figure}

In Fig. 1, we also overplot a theoretical AMR
derived from a simple closed-box chemical enrichment model.  This
model assumes that the stellar system was completely gaseous 13.5
Gyrs ago, and converted gas into stars at a constant rate (set so
that the gas supply has just been depleted at the present
day). We assume Z$_{\odot}$ = 0.02. We find that 
using a yield of 0.004 gives a reasonable fit to the data.

The model AMR provides a very good representation of the chemical
enrichment history of the Sgr dwarf over the last $\sim$13 Gyrs. 
This supports the suggestion of Layden \& Sarajedini (2000) that the
Sgr dwarf may have formed stars and GCs without significant
infall or explusion of gas. 
Simple closed-box models are also 
inferred for the Fornax dwarf (Pont \etal~2003) and the LMC/SMC
galaxies (Piatti \etal~2002), but more 
complicated enrichment models are also possible
(e.g. Smecker-Hane \& McWilliams 2002).
We now proceed to compare the Sgr
model AMR with cluster and field star data for the Canis Major
dwarf.

\section{The Age-Metallicity Relation of the Canis Major Dwarf Galaxy}

As noted in the introduction, Martin \etal (2003) identified four
GCs associated with the Canis Major dwarf and Bellazzini \etal (2003c) added
two old open clusters. The open clusters, AM-2 and Tom~2, have
approximate ages and metallicities of 5 Gyrs and [Fe/H] =
$-0.5$ (Lee 1997) and 4 Gyrs and [Fe/H] = $-0.45$ 
(Brown \etal~1996) respectively. 

\begin{figure}
\plotone{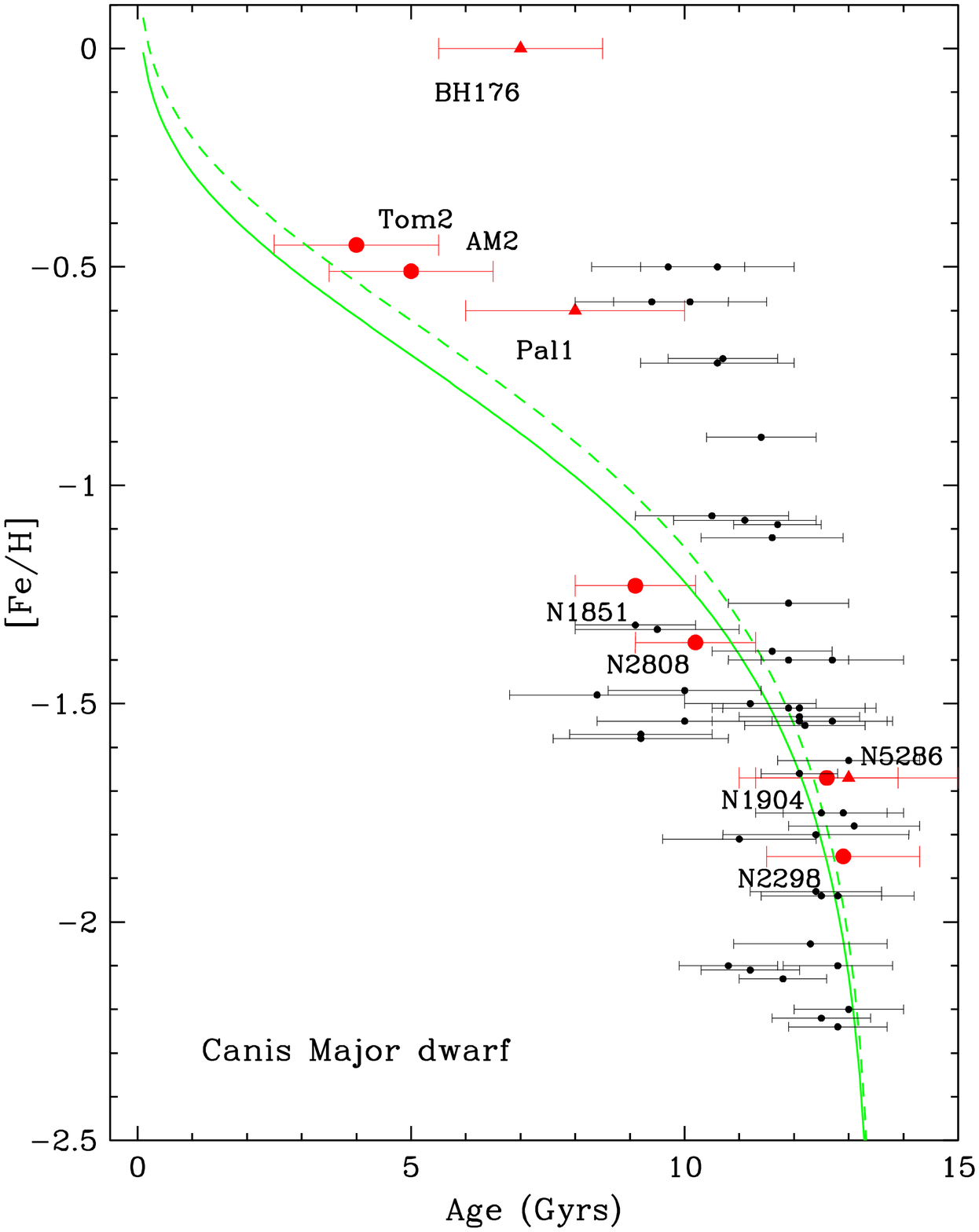}
\caption[forbes.f2.ps]{Age-metallicity relation (AMR) for the Canis Major
dwarf galaxy. The four globular clusters (GCs) and two 
open clusters are shown by
filled circles. Three Monoceros ring GCs are shown by
filled triangles. The typical error in GC [Fe/H] is $\pm$ 0.1 dex.
The solid curve shows the Sgr galaxy AMR 
from Fig. 1, and the dashed line a 20\% higher yield.
The small filled circles show the remaining Milky Way GCs. 
The GC BH~176 is not
consistent with the AMR of the Canis Major galaxy, 
and therefore unlikely to be a former member. }
\end{figure}

In Fig. 2 we show these six Canis Major clusters compared
to the Sgr model AMR. Given the similar inferred
luminosities for the two galaxies (Martin \etal
2003), we might expect the Sgr AMR to be a reasonable
approximation for Canis Major. Indeed this appears to be the
case. Both galaxies are consistent with a simple closed-box chemical enrichment
model, although the stellar yield in the Canis Major galaxy may have been
slightly higher than in Sgr.

We also show the three additional Mono ring GCs candidates suggested by
Crane \etal (2003). The GCs Pal~1 and NGC~5286 are generally consistent
with the high yield AMR, but the very young GC BH~176 deviates
significantly ($>$3$\sigma$) from it. We conclude that BH~176 is
{\it unlikely} to be a former member of the Canis Major dwarf galaxy. 

 

Fig. 2 also shows the ages and
metallicities of the remaining (i.e. non Sgr and non Canis Major) Milky Way
GCs in the list of Salaris \& Weiss (2002). The model AMR
deviates from the Milky Way GC distribution at intermediate-to-high metallicities. 
The Milky Way GCs reveal a much steeper AMR (i.e., they are more
chemically enriched at a given age). A closed-box model would
require a much higher yield to achieve such an AMR. However,
detailed studies of the Milky Way AMR indicates that such a model
is a poor representation of reality, and additional processes are
required, such as pre-enrichment or infall (e.g., Tinsley 1980).

Some of the metal-poor Milky Way GCs deviate from the mean trend
line to younger ages. These GCs typically lie at large
galactocentric radii 
and have been classified as ``young halo'' GCs (Zinn 1993). These GCs
include: Eridanus (age = 8.4 Gyrs, [Fe/H] = $-1.48$, R$_{GC}$ =
95.2 kpc), Pal 3 (9.2, $-1.57$, 95.9), Pal 4 (9.2, $-1.58$, 111.8), and Rup 106
(10.4, $-1.90$, 18.5). These, and other GCs with large
galactocentric radii, have been suggested as prime candidates for
accreted GCs by van den Bergh (2000). Thus, Fig. 2 provides
further circumstantial evidence that these GCs may have been
accreted from tidally captured galaxies. If these `young halo' GCs
and the Canis Major and Sgr GCs are excluded from a Milky Way
analysis, then the age spread of the Salaris \& Weiss (2002)
sample of metal-poor GCs is reduced from $\sim 3-4$ Gyrs to $\la
1$ Gyr. This would make any ELS-type collapse (Eggen, Lyden-Bell
\& Sandage 1962) very rapid indeed.

\section{The Size-Galactocentric Distance Relation}

The existence of a size-galactocentric distance
relation for Milky Way GCs suggests that the bulk of the system
formed \emph{in situ} (van den Bergh 1995). This point holds because
the half-mass radius of a GC (unlike the core and tidal radii) is
expected to be largely unaffected by internal or external
dynamical processes over a Hubble time (Murray \& Lin 1992), and
any compact GCs at large galactocentric distances  are likely to have
survived to the present day (Ashman \& Zepf 1998).
In Fig. 3 we show the
half-mass radius versus galactocentric distance for the Canis
Major GCs compared to other Milky Way GCs. Both quantities are
taken from Harris (1996). The Canis Major GCs have relatively
small half-mass radii for their galactocentric
distance. This lends further support to their origin
outside of the Milky Way. We note, however, that the Sgr GCs span a 
large range in half-mass sizes. This suggests that a
small range in GC sizes, while suggestive, is not a common feature
of all dwarf GC systems.

\begin{figure}
\plotone{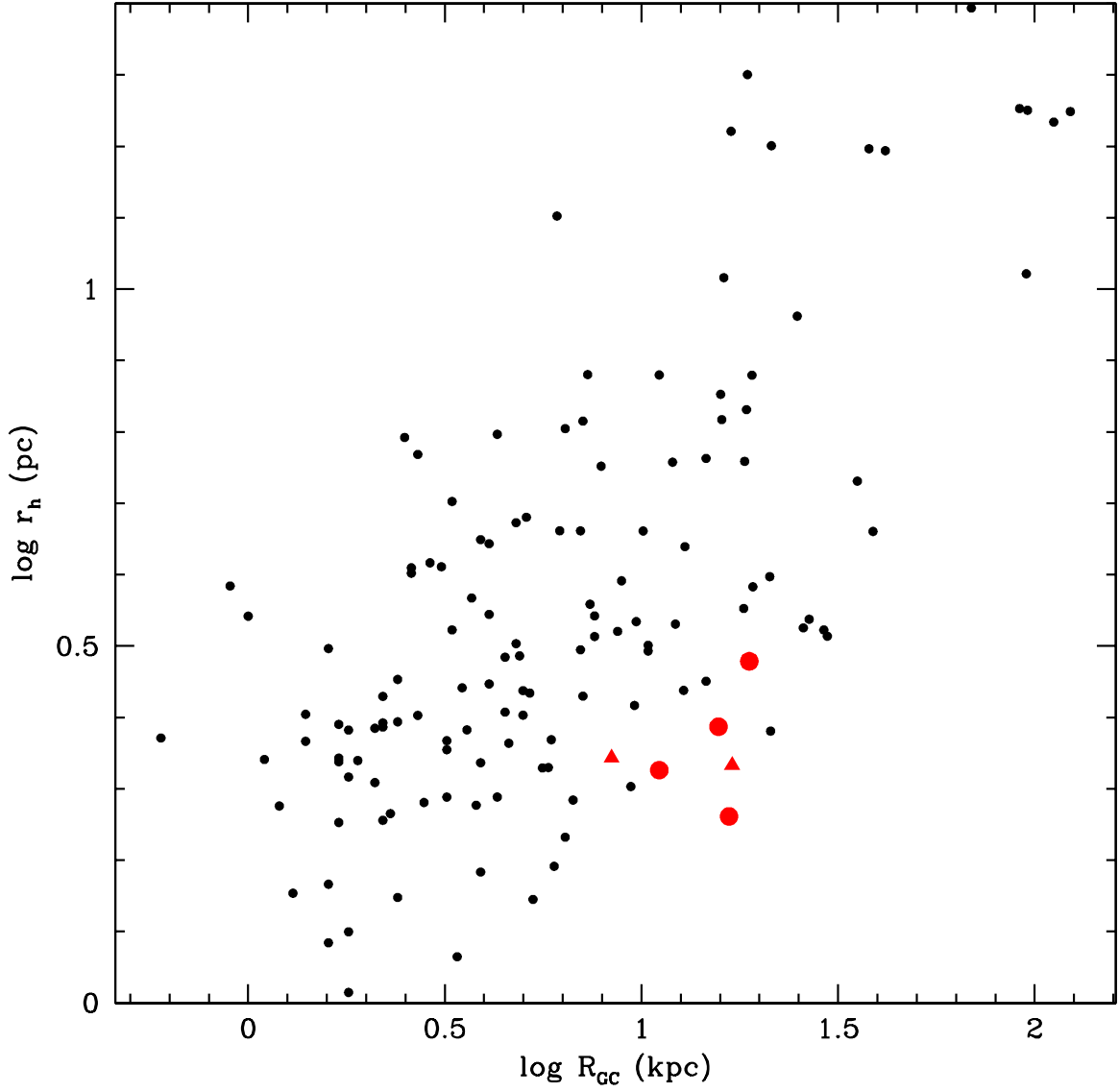}
\caption[forbes.f3.ps]{Globular cluster half-mass radii r$_h$
vs.~galactocentric distance R$_{GC}$. The plot shows Milky Way GCs 
(small filled circles), Canis Major (large filled circles) and
the Mono ring GCs Pal~1 and NGC~5286 (filled triangles). Typical
errors are $\pm$ 0.1 pc in the half-mass radius and $\pm$ 0.1 kpc in
distance. The Canis Major/Mono ring 
globular clusters are on average more compact than Milky Way
GCs. 
}
\end{figure}



\section{The Contribution of Accreted Dwarf Galaxies}

It is likely that the Sgr and Canis Major dwarfs are not the only
satellites accreted by our Galaxy. 
There are good examples in other disk galaxies of {\it accreting}
(Forbes \etal~2003) and {\it accreted} (Ibata \etal~2001)
satellite galaxies that contribute to the build-up of galactic
halos. For M31, gas associated with an accreted satellite may
have induced some new GC formation (Beasley \etal~2004).
How significant are these accretions in the growth of a typical
spiral galaxy and to which galactic component do they contribute ?

Martin \etal~(2003) suggested that the Canis Major
dwarf is a major building block of the Milky Way's thick
disk. This interpretation fits in well with the simulations of
Abadi \etal~(2003), in which thick disks are the result of
satellite accretions in the galactic plane. The thick disk GC
system is traditionally thought to be a flattened, highly
rotating system with a relatively high mean metallicity (Zinn
1985). Today the metal-rich GC system is often considered to be
associated with the bulge of our Galaxy (Minniti 1995), and
similarly for external galaxies (Forbes, Brodie \& Larsen 2001).
However, the mean metallicity of the four Canis Major GCs
identified by Martin \etal (2003) is
[Fe/H] = $-1.5$, i.e. consistent with the mean for the 
{\it metal-poor} GC system. 
Even the highest metallicity GC, NGC~1851, is
still relatively metal-poor with [Fe/H] = $-1.23$. Such
metallicities are not typical of thick disk/bulge GCs but are
generally associated with halo GCs. Futhermore, only one 
Sgr GC (Pal~12), and none of the five Fornax dwarf GCs, have [Fe/H] $\ge
-1$. This suggests that even when satellite accretions do add to
the GC system of the Milky Way, such additions are unlikely to
contribute significantly to a thick disk or bulge population of GCs.
We note also that the alpha-element ratios of thick disk stars 
(Feltzing \etal 2003) are generally super-solar and so {\it not}
consistent with (solar ratio) stars found in Local Group dwarfs such as
the Sgr dwarf (Bonifacio \etal 2004). Such solar abundance ratios indicate chemical
enrichment over an extended time period, as suggested by the AMRs
for Sgr and Canis Major (see Figures 1 and 2).

Several studies have placed limits on the importance of
dwarf galaxy accretion to the Milky Way's halo.
Based on studies of halo
stars, van den Bergh (2000) has argued that 3--7 Sgr-like dwarf
accretions may have occurred over the Milky Way's lifetime. 
By comparing halo star ages with those from dwarf galaxies, Unavane,
Wyse \& Gilmore (1996) have argued that only $\sim 10\%$ of the
\emph{halo} mass could have come from dwarfs.  
Gilmore \& Wyse (1998) went on to examine the orbits of halo stars
with alpha-element ratios that are similar to dwarf galaxy stars
and concluded they were unlikely to have come from accreted
dwarfs. 

\section{Conclusions}

The identification of globular clusters (GCs) with the accreted
dwarf galaxies Sgr
and Canis Major has highlighted the fact that GC properties such as
the horizontal branch morphology, prograde or retrograde orbits, 
range in cluster magnitudes, metallicity and galactocentric
distance do not provide a unique signature of an accreted GC. 

We have examined the relation between half-mass size and
galactocentric distance for the Canis Major GC system, finding
that all of the associated GCs have smaller than average
sizes. This suggests an origin outside of the Milky Way.

We have derived a simple closed-box age-metallicity relation
that provides a good representation of the chemical enrichment
history of the Sgr and Canis Major dwarf galaxies. This is
consistent with formation of these GCs outside of the Milky Way. 
The model AMR deviates from the Milky Way GC distribution at
intermediate-to-high metallicities, thus providing an alternative
and fairly robust method of identifying
accreted metal-rich GCs within the Milky Way GC system. 
Based on
the literature age and metallicity measurements for the GC BH~176, we argue it
is unlikely to be a former member of the Canis Major
galaxy. 

We 
support earlier suggestions that the younger, metal-poor Milky
Way GCs are prime candidates for accreted GCs from as yet
unidentified galaxies. This would reduce
the GC age spread, at a given metallicity, to less than a Gyr and
imply that any halo collapse was very rapid. 
As the age-metallicity relation of the known accreted GCs is
distinct from that of other 
Milky Way GCs for [Fe/H] $\ge -1.3$, it 
suggests that accretion was not a major factor in building the
stellar mass of the thick disk or bulge. 
The majority of the Milky Way GC system and, by implication, the
Galaxy itself formed {\it in situ}.

\acknowledgements

This work was supported by NSF grant number AST-0206139 and an
NSF Graduate Research Fellowship to JS. We thank R. Proctor for
his helpful comments.

\end{document}